# ALICE: The Ultraviolet Imaging Spectrograph aboard the New Horizons Pluto-Kuiper Belt Mission

S. Alan Stern[a], David C. Slater[b], John Scherrer[b], John Stone[b], Greg Dirks[b], Maarten Versteeg[b], Michael Davis[b], G. R. Gladstone[b], Joel Wm. Parker[a], Leslie A. Young[a], and O. H. W. Siegmund[c]

[a]Southwest Research Institute, 1050 Walnut St., Suite 400, Boulder, CO 80302
[b]Southwest Research Institute, 6220 Culebra Rd., San Antonio, TX 78238
[c]Sensor Sciences, 3333 Vincent Road, Pleasant Hill, CA 94523

## ABSTRACT

The New Horizons ALICE instrument is a lightweight (4.4 kg), low-power (4.4 Watt) imaging spectrograph aboard the New Horizons mission to Pluto/Charon and the Kuiper Belt. Its primary job is to determine the relative abundances of various species in Pluto's atmosphere. ALICE will also be used to search for an atmosphere around Pluto's moon, Charon, as well as the Kuiper Belt Objects (KBOs) that New Horizons hopes to fly by after Pluto-Charon, and it will make UV surface reflectivity measurements of all of these bodies as well. The instrument incorporates an off-axis telescope feeding a Rowland-circle spectrograph with a 520-1870 Å spectral passband, a spectral point spread function of 3-6 Å FWHM, and an instantaneous spatial field-of-view that is 6 degrees long. Different input apertures that feed the telescope allow for both airglow and solar occultation observations during the mission. The focal plane detector is an imaging microchannel plate (MCP) double delay-line detector with dual solar-blind opaque photocathodes (KBr and CsI) and a focal surface that matches the instrument's 15-cm diameter Rowland-circle. In what follows, we describe the instrument in greater detail, including descriptions of its ground calibration and initial in flight performance.

## 1. BACKGROUND

New Horizons (NH) is the first mission in NASA's New Frontiers line of mid-scale planetary exploration missions. More specifically, New Horizons is a scientific reconnaissance mission to the Pluto-Charon system in 2015 and possibly one or more Kuiper-Belt Objects (KBOs) thereafter; en route, the mission includes a Jupiter Gravity Assist (JGA) flyby in 2007 that also includes extensive scientific observations. NH launched on 19 January 2006.

New Horizons ALICE (also referred to as P-ALICE, for Pluto-ALICE) is a compact, low-cost instrument designed to perform spectroscopic investigations of planetary atmospheres and surfaces at extreme (EUV) and far-ultraviolet (FUV) wavelengths between 520 and 1870 Å. It is a direct derivative of the Pluto mission "HIPPS" UV spectrograph (HIPPS/UVSC), developed at Southwest Research Institute (SwRI) in the mid-1990s with funds from NASA, JPL, and SwRI[1]. Later, the HIPPS/UVSC instrument was re-optimized for the European Space Agency (ESA) Rosetta comet orbiter mission by increasing its sensitivity, instantaneous field-of-view (FOV), and wavelength coverage, and by adding a lightweight microprocessor. This instrument, called Rosetta-ALICE (R-ALICE)[2], was successfully launched on the Rosetta spacecraft in March 2004, and is operating successfully on a 10-year mission to rendezvous with and orbit about comet 67P/Churyumov-Gerasimenko. For NH, ALICE improvements include a solar occultation channel (SOC) for direct atmospheric solar occultation observations, various reliability enhancements, and a slightly different passband than R-ALICE. Extensive information on the New Horizons mission and its scientific objectives can be found in the articles in this volume by Stern et al., Young et al., and Fountain et al.

## 2. SCIENTIFIC OBJECTIVES

As described in Young et al. (2007) in this volume, one of the three primary (i.e., Group 1) NH mission objectives is to "Characterize the neutral atmosphere of Pluto and its escape rate." P-ALICE is specifically designed to address this measurement objective. Moreover, P-ALICE addresses both Group 1 and Group 2 mission science objectives at Pluto, including important questions about Pluto's atmosphere, such as determining:

- The mixing ratios of $N_2$, CO, $CH_4$ and noble gases.
- The vertical density and temperature structure (e.g., gradient) of the upper atmosphere.
- The atmospheric haze optical depth.
- The atmospheric escape rate and escape regime.

P-ALICE will also address New Horizons objectives concerning the search for an atmosphere around Charon and it will study the UV reflectance properties of the surfaces of Pluto, Charon, Pluto's small moons, and KBO targets. Searches for atmospheres around KBOs will also be performed.

As an illustration of expected scientific performance, Figure 1 shows results of modeling of solar occultation and airglow observations by P-ALICE. $N_2$ continuum and band absorption structure at wavelengths <100 nm dominates Pluto's predicted EUV opacity, allowing sampling of the uppermost atmosphere. From 100-150 nm, $CH_4$ dominates the opacity, providing a window on the middle atmosphere from roughly 300 to 1200 km. At wavelengths >150 nm, longer hydrocarbons with strong FUV absorption bands, such as $C_2H_2$ and $C_4H_2$, along with hazes, are expected to provide information about the lowest ~100 km of Pluto's atmosphere. Figure 1 also shows predicted airglow spectra at Pluto indicating the potential feasibility of the detection of atomic species like H, N, and Ar, the $N^+$ ion, as well as molecular CO and $N_2$.

Of course, the actual ability of P-ALICE to detect and study Pluto's atmosphere depends sensitively on the degree of atmospheric collapse that might occur on Pluto (e.g., McNutt et al. 1997; Spencer et al. 1997; Trafton et al. 1997) by the time of NH's 2015.5 encounter.

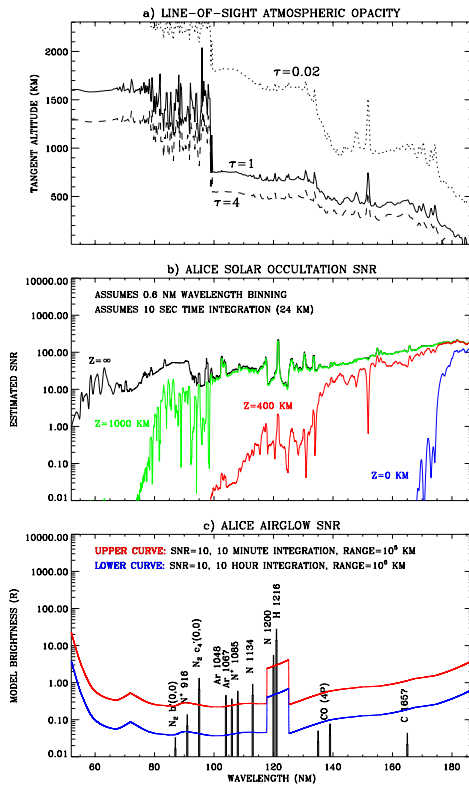

**Figure 1.** a) Limb-viewing, line-of-sight opacity for model Pluto as a function of wavelength and tangent altitude, for several values of optical depth τ; b) Solar occultation NR estimates based on the line-of-sight transmitted solar flux through Pluto's atmosphere, for several tangent altitudes, as a function of wavelength for integration times of 10 sec; c) Model dayglow brightnesses for emissions resulting from photoelectron impact excitation, dissociative excitation, and resonant scattering of sunlight. Here the upper and lower curves show expected P-ALICE SNR~10 for 10-min and 2-hr integrations at a range of $10^5$ km.

## 3. INSTRUMENT DESCRIPTION

### 3.1. Overview

As shown in Figure 2, the P-ALICE UV spectrograph is comprised of a telescope, a Rowland-circle spectrograph, a detector at the focal plane, and associated electronics and mechanisms. Figure 3 shows a photograph of the exterior of the P-ALICE flight model.

P-ALICE has two separate entrance apertures that feed light to the telescope section of the instrument: the airglow channel (AGC) aperture, and the SOC aperture. The AGC's 40x40 mm$^2$ entrance aperture is the front end of the instrument; the SOC aperture is a small 1-mm diameter opening located perpendicular to the side of the telescope section of the instrument (see Fig. 2). The SOC stops down the entrance area relative to the AGC by a factor of ~6400 to allow for solar occultation studies without detector saturation. The SOC is orthogonal to the AGC to boresight it with the REX Radio Science Experiment on NH for the nearly simultaneous solar and Earth occultations that occur in the Pluto system. A small relay mirror reflects the SOC beam into the telescope optical path. The SOC and AGC may also be used for stellar occultations of flyby targets.

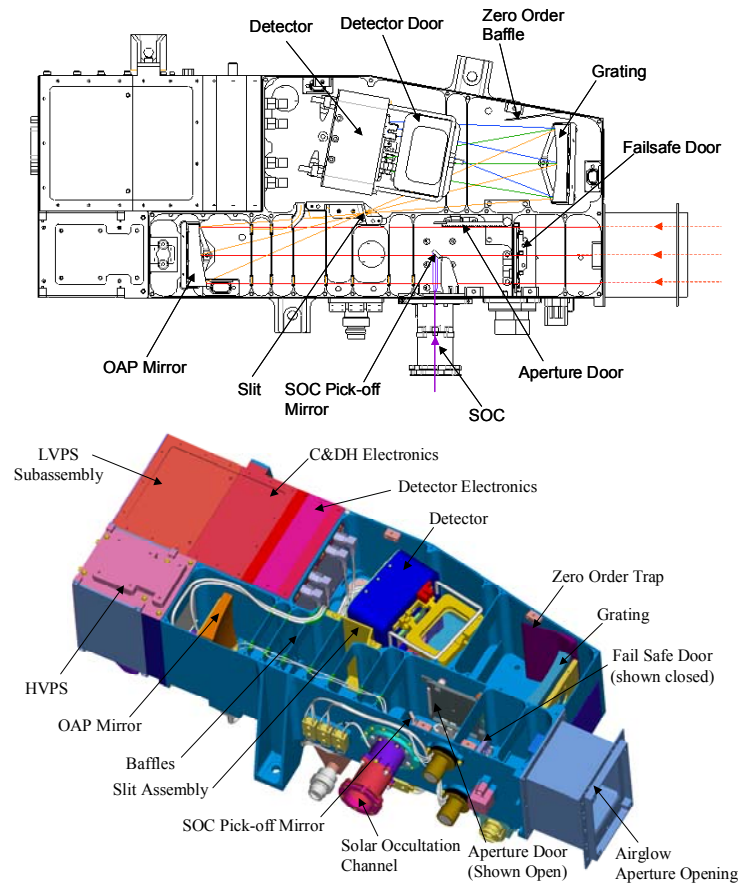

**Figure 2.** (Top) An opto-mechanical drawing of P-ALICE with central-axis light rays shown. (Bottom) 3-D opto-mechanical view. ALICE is approximately 45 cm long.

Light entering either aperture is collected and focused by an f/3 off-axis paraboloidal (OAP) primary mirror at the back end of the telescope section onto the instrument's entrance slit. After passing through the entrance slit, the light falls onto a toroidal holographic diffraction grating, which disperses the light onto a double-delay line (DDL) microchannel plate (MCP) detector[6]. The 2-D format MCP detector uses side-by-side, solar-blind photocathodes—potassium bromide (KBr) and cesium iodide (CsI)—and has a cylindrical, curved MCP-stack that matches the Rowland-circle. P-ALICE is controlled by an Intel 8052 compatible microcontroller, and utilizes lightweight, compact, surface-mount electronics to support the science detector, as well as redundant power supplies, command and data handling, instrument support, and other spacecraft interface electronics.

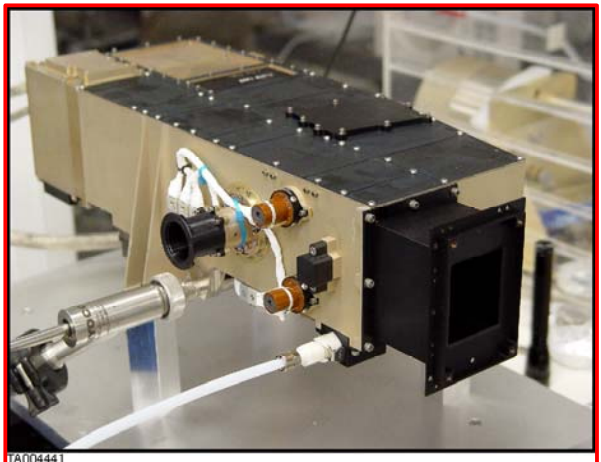

**Figure 3.** Photograph of P-ALICE.

### 3.2. Opto-mechanical Design Overview

P-ALICE's telescope mirror and diffraction grating are both constructed from monolithic pieces of aluminum, coated with electroless nickel and polished using low-scatter polishing techniques. These aluminum optics, in conjunction with the aluminum housing, form an athermal optical design.

The P-ALICE optics are over-coated with sputtered SiC for optimum reflectivity within the instrument's spectral passband[7]. Additionally, P-ALICE's zero order baffle light trap is treated with a nickel-phosphorus (Ni-P) black coating with very low surface reflectance at EUV/FUV wavelengths[10]. Further still, additional control of internal stray light is achieved via (i) UV-absorbing internal baffle vanes within the telescope and spectrograph sections of the housing, (ii) a holographic diffraction grating that has low scatter and near-zero line ghost problems, and (iii) alodyned internal surfaces[8,9].

The spectrograph entrance slit design is shown in Figure 4. It has two contiguous sections: a narrow, rectangular slit segment with a field-of-view of 0.1ºx4.0º; and a fat, square slit segment with a 2.0ºx2.0º field-of-view. The large, 2.0ºx2.0º slit opening is designed to ensure that the Sun is captured within the instrument's field-of-view during the solar occultation observations of Pluto and Charon that must occur nearly simultaneously with the radio science occultation observations performed with the REX instrument that uses the spacecraft's high gain antenna (HGA). During solar occultation observations, the HGA would be pointed back at the Earth, with P-ALICE pointed back at the Sun. P-ALICE is aligned on the spacecraft with a 2º tilt along the spectrograph's spatial axis so that the Sun is centered within the SOC square FOV when the HGA is pointed at the Sun (to improve the quality of the radio occultation data, the flyby is planned to occur when Pluto is near opposition as seen from Earth, i.e., around mid-July). A misalignment up to ±0.9º between the SOC FOV and the HGA boresight is accommodated by the large SOC FOV. The spectral PSF of the SOC is ~5 Å across the passband[3].

Heaters are mounted to the back surfaces of the OAP mirror, SOC relay mirror bracket, and the grating to prevent condensation of contaminants onto the optics during flight. To protect the sensitive photocathodes and MCP surfaces from exposure to moisture and other harmful contaminants during ground operations, instrument integration, and the early stages of the mission, the detector tube body assembly is enclosed in a vacuum chamber with a front hermetic door that includes a magnesium fluoride (MgF$_2$) UV-transparent window that was permanently opened during the commissioning phase of the flight. The front aperture door can be commanded closed for additional protection of the optics and detector from direct solar exposure during propulsive maneuvers.

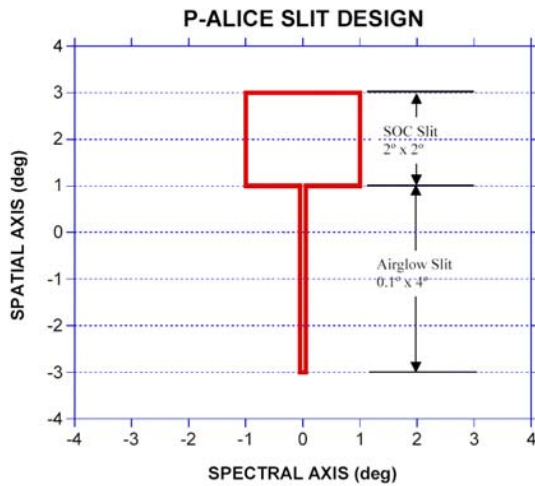

**Figure 4.** The P-ALICE entrance slit design. The narrow 0.1ºx4º slit is sometimes designated the "airglow" slit; the larger 2ºx2º opening above the airglow slit is sometimes called the SOC slit. The airglow optical boresight is centered at (0º, 0º).

P-ALICE's aperture door is its sole multiple-use actuator, and can be opened and closed via command; it is qualified for 10,000 cycles, but is expected to be used <1,000 times on New Horizons. A limited angle torque (LAT) motor with a direct tie to the aperture door hinge pin provides the torque required to open and close the door. The aperture door remained closed until it was determined that the spacecraft had sufficient time to fully outgas (i.e., ~50 days) after launch. The aperture door was then opened to allow residual outgassing of the interior of the instrument and for heater decontamination sessions of the optics.

Following these activities, the detector door, which uses a Wax Pellet Actuator (WPA) was also opened; this was followed by a slow, controlled, high voltage ramp up of the detector MCP stack to begin detector check out. The detector door design is an improved version of that flying aboard the R-ALICE instrument[2]. During ground operations the door was opened numerous times and manually reset; during flight, once open, the door need not and cannot be re-closed. A MgF$_2$ window provides a method to get light to the detector during ground ops when the detector door is closed; this window also provided protection to science objectives at wavelengths longer than 120 nm in the event the detector door had not opened in flight.

A one-time opening failsafe aperture door is available on the aperture door in the event that it ever becomes stuck in the closed state. With the failsafe door open, the AGC sensitivity is ~10% that of the fully opened AGC with the aperture door in the open state. A larger failsafe aperture was not considered prudent owing to the long cruise exposure (up to 9+ years) that could risk UV photolysis of hydrocarbons on the optics. The SOC pinhole aperture cover will be opened shortly before the Jupiter flyby when the risk of accidental SOC aperture Sun pointing that can cause UV photolysis of the optics has been sufficiently reduced owing to the spacecraft heliocentric distance.

### 3.3. Detector and Detector Electronics Overview

P-ALICE's detector utilizes an MCP Z-stack that feeds its DDL readout array[6]. To cover the 520-1870 Å passband and 6° spatial FOV, the size of the detector's active area is 35 mm (in the dispersion direction) x 20 mm (in the spatial dimension), with a pixel format of (1024x32)-pixels. The 6° slit-height is imaged onto the central 22 of the detector's 32 spatial channels; the remaining spatial channels are used for dark count monitoring. The pixel format allows Nyquist sampling with a spectral resolution of 3.6 Å, and a spatial resolution of ~0.6°. The input surface of the Z-stack is coated with opaque photocathodes of KBr (520-1180 Å) and CsI (1250-1870 Å)[11], in two different wavelength regions.

The MCP Z-stack is composed of three 80:1 length-to-diameter (L/D) MCPs that are cylindrically curved with a radius-of-curvature of 75 mm to match the Rowland-circle for optimum focus; the total Z-Stack resistance at room temperature is ~300 MΩ. The MCPs are rectangular in format (46x30 mm$^2$), with 12-µm diameter pores. Above the MCP Z-Stack is a repeller grid that is biased ~900 V more negative than the top of the MCP Z-Stack. This repeller grid reflects electrons liberated in the interstitial regions of the MCP back down to the MCP input surface to enhance the detector quantum efficiency. The MCP Z-stack requires a high negative voltage bias of ~ –3 kV; an additional –600 V is required between the MCP Z-stack output and the anode array (the anode array is referenced to ground). The intrinsic dark count rate of the flight MCP stack is quite low—less than 3 counts sec$^{-1}$ over its entire active area[3]; however, in flight radiation (both natural and from the NH RTG) has raised the background count rate to almost 100 Hz.

To prevent saturation effects on the detector electronics during observations, it is necessary to attenuate the solar Lyman-α emission brightness to an acceptable count rate level well below the maximum count rate capability of the electronics (i.e., below $3\times10^4$ c sec$^{-1}$). An attenuation factor of at least an order of magnitude is required to achieve this lower count rate. This is easily achieved by physically masking the MCP active area around Lyα during photocathode coating so it remains bare. The bare MCP glass has a quantum efficiency about 10 times less than that of KBr at 1216 Å. This masking technique has been successfully demonstrated in the past with the DDL detector aboard the Rosetta-ALICE instrument[2] and the SUMER instrument on SOHO[12].

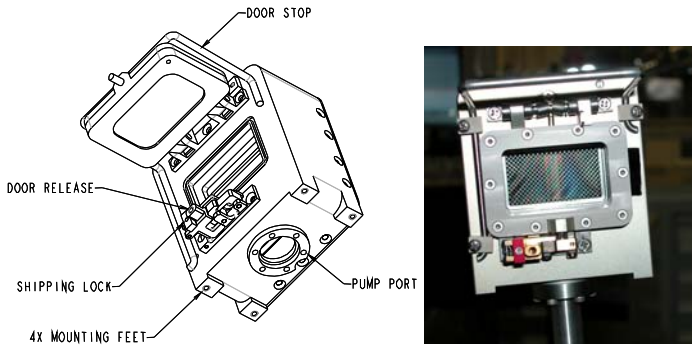

**Figure 5.** (Left) 3-D schematic of DDL detector vacuum housing with the detector door shown in the open position. (Right) Photograph of the flight DDL detector/vacuum housing. The repeller mesh and MCP Z-stack are visible through the MgF$_2$ window.

Surrounding the detector tube body is a vacuum chamber housing made of aluminum and stainless steel (Figure 5). As described above, this vacuum chamber is designed to protect the MCP stack and the KBr and CsI photocathodes against damage from moisture exposure during ground handling and from outgassing constituents during the early stages of flight. It also allows the detector to remain under vacuum during ground operations, testing, handling, and transportation. The MCP/anode tube body assembly mounts to the rear stainless steel vacuum flange. This flange mates to the aluminum vacuum housing with a vacuum tight O-ring seal, and contains two high-voltage (HV) feedthrough connectors that are welded to the flange (HV input and return), as well as four welded microdot feedthrough connectors for the four analog signal outputs from the DDL anode.

The DDL detector's electronics include preamplifier circuitry, time/charge digital converter circuitry (TDC), and pulse-pair analyzer circuitry (PPA) cards. These three boards are mounted inside a separate enclosed aluminum housing that mounts to the rear of the spectrograph section (just behind the detector itself). The detector electronics require ±5 VDC, and draw ~1.1 W.

The detector electronics amplify and convert the detected output pulses from the MCP Z-stack to pixel address locations. Only those analog pulses output from the MCP that have amplitudes above a set threshold level are processed and converted to pixel address locations. For each detected and processed event, a 10-bit x (spectral) address and a 5-bit y (spatial) address are generated by the detector electronics and sent to the P-ALICE C&DH electronics for data storage and manipulation. In addition to the pixel address words, the detector electronics also digitizes the analog amplitude of each detected event output by the preamplifiers and sends this data to the C&DH electronics. Histogramming of this "pulse-height" data creates a pulse-height distribution used to monitor the health and status of the detector during operation.

An analog count rate signal is output from the detector electronics to the C&DH to allow monitoring and recording of the detector total array count rate. This count rate data is updated once per second and is included in the instrument housekeeping (HK) data.

A built-in "stim-pulser" is included in the electronics that simulates photon events at two pixel locations on the array (located in the upper right and lower left corners of the active array). This pulser can be turned on and off by command and allows testing of the entire detector and C&DH electronics signal path without having to power on the detector high-voltage power supplies or put light on the detector.

### 3.4. Instrument Electrical Design Overview

The P-ALICE instrument support electronics include the redundant low-voltage power supplies (LVPS), actuator electronics, C&DH electronics, optics decontamination heaters, and redundant detector high-voltage power supplies (HVPSs).

A simplified block diagram of these electronic subsystems is shown in Figure 6; Figure 7 contains a more detailed block diagram. These sub-systems are controlled by a radiation-hardened version of the Intel 8052 microprocessor with 32 kB of fuse programmable PROM, 128 kB of EEPROM, 32 kB of SRAM, and 128 kB of acquisition memory. The C&DH electronics are contained on four circuit boards located just behind the detector electronics (see Figure 2). We now describe each major block.

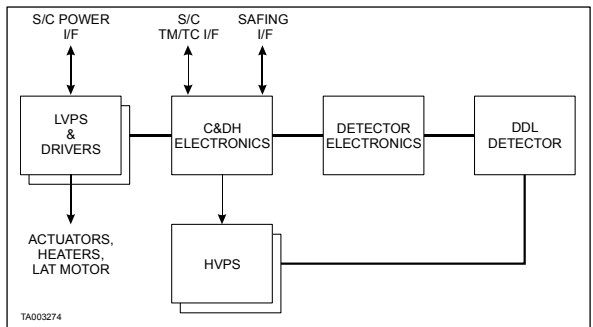

**Figure 6.** Simplified electronics block diagram.

<u>Low-Voltage Power Supply Electronics.</u> LVPS electronics and associated drivers are redundant, and are each composed of DC/DC converters designed to convert the +30V supplied spacecraft power to ± 5 VDC (power ~4 W) and +2.5 VDC (power ~0.4 W) to the ALICE electronics. Five boards make up the LVPS: two LVPS boards (both primary and redundant), an EMI filter board, a heater/actuator board, and a motherboard. The LVPS electronics interfaces to the spacecraft via three links: two links to the +30V spacecraft power bus for (i) instrument power and (ii) actuator/heater power; the third link is a safing connector interface that prevents inadvertent HVPS and/or actuator (i.e., launch latch, detector door, fail-safe door, SOC door, and/or detector vent valve) operations. The switching circuit for the decontamination heaters and the motor controller that operates the front aperture door is in the LVPS electronics.

<u>Command–and–Data–Handling Electronics.</u> The ALICE C&DH electronics are controlled by the 8052 microprocessor, and provide the command interface as well as the science and telemetry interfaces with the NH spacecraft. They handle the following instrument functions: (i) interpretation and execution of commands to the instrument, (ii) collection of raw event data from the detector to the dual-port acquisition memory (see Fig. 7) in either the pixel list or histogram data collection modes (described in the following section), (iii) telemetry formatting of science and housekeeping data, (iv) control of the redundant detector HVPSs, (v) operation of all actuator and heater functions, (vi) control of the front aperture door LAT motor, (vii) control of the housekeeping ADCs used to convert analog housekeeping data to digital data for inclusion into the TM data stream, and (viii) monitoring of detector health and status via the detector analog count rate, and the HVPSs' high voltage and MCP strip current values. For P-ALICE flight operations, the Intel 8052 microprocessor operates with a clock frequency of 4.0 MHz. P-ALICE low-speed HK telemetry and telecommand interface is accomplished via an RS-422 asynchronous interface to the NH spacecraft. A second FPGA within the C&DH electronics implements a UART that sends the low-speed data to the spacecraft.

Decontamination Heaters. As described earlier, a redundant decontamination heater (~1 W resistive) is bonded to the backside surface of both the OAP mirror and the grating substrates. Along with each heater, two redundant thermistors are also mounted to the back of each substrate to monitor and provide control feedback to the heaters. The C&DH electronics can separately control each heater.

High Voltage Power Supplies. Redundant HVPSs are located in a separate enclosed bay behind the OAP primary mirror (see Figure 2). The HV output of these two supplies are diode-or'ed together to the single HV terminal on the MCP/anode assembly. Each supply can output a maximum of –6 kV to the top of the MCP/repeller grid assembly—well above the nominal operational HV level of –4.5 kV. The voltage between the MCP output and the anode array is fixed at –600 V using Zener diodes between the output and ground. The voltage to the MCP Z-stack is fully programmable by command between 0 and –6 kV; they require ±5 VDC and consume ~0.3 W.

### 3.5. Data Collection Modes Overview

P-ALICE has two detector data collection modes described in detail below: (i) pixel list mode; and (ii) histogram mode. The P-ALICE flight software controls both of these modes. Data is collected in a dual port acquisition memory that consists of two separate 32kx16-bit memory channels.

In Pixel List Mode (PLM) a continuous time series of photon counts is generated with a series of interspersed time hacks at a programmable interval, which can be as short as 4 msec. In PLM, each memory channel can hold up to 32k detector and/or time-hack events; where each detector event consists of a 16-bit word—either an x-position word 10 bits in length, a y-position word 5 bits in length, and a single status bit that distinguishes between a time hack word and a detector event word; or a time hack word that includes a single status bit plus 15-bits that encodes the instrument on-time. When 32k detector address and time-hack events have accumulated in one of the two acquisition memories, that acquisition memory stops accumulating event data and begins to read the data out to a parallel-to-serial converter and a low-voltage differential signaling (LVDS) interface to the spacecraft on-board memory. At the same time that this LVDS readout is taking place, the other side of the dual acquisition memory continues to collect detector and time-hack data until it fills up, whereupon it reads out to the LVDS interface to the spacecraft memory, while the first acquisition memory takes over collecting detector and time-hack data. This back-and-forth data collection flow between both acquisition memories is called "ping-pong" acquisition—it allows contiguous readout of detector event data as long as the data event rate does not exceed the rate at which the data can be read out of memory to the LVDS interface. The "ping-pong" acquisition process is controlled using logic encoded in one of two FPGAs within the C&DH electronics.

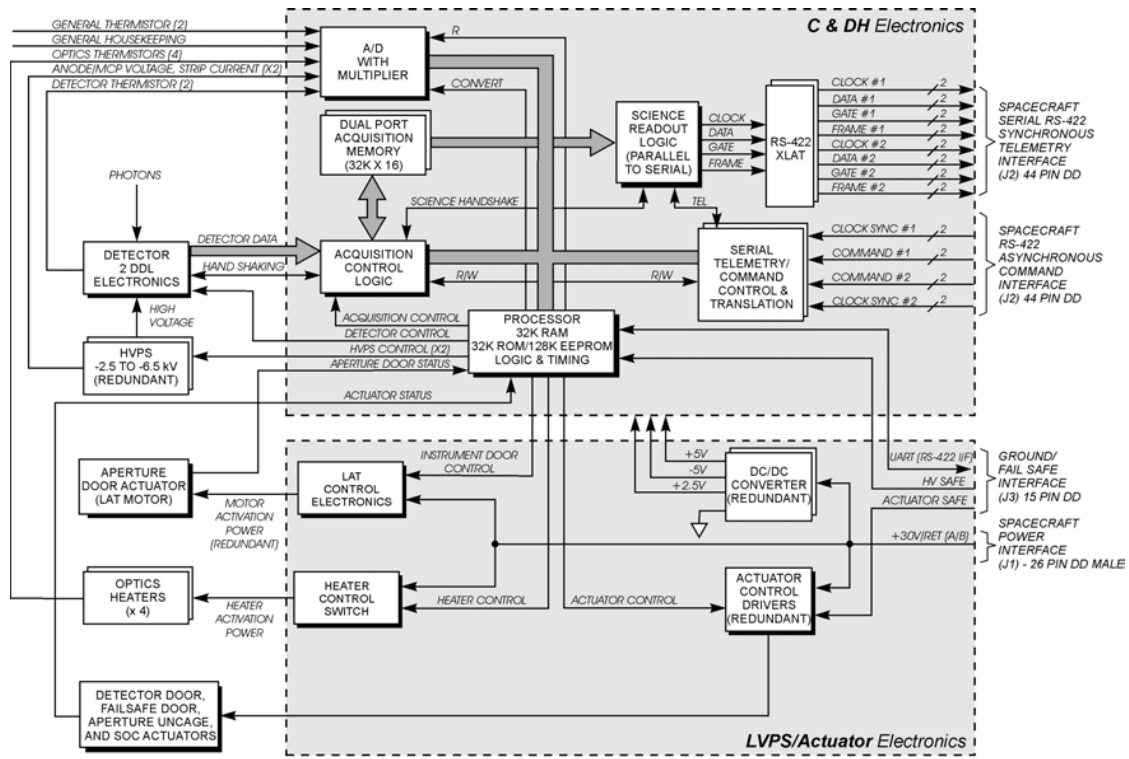

**Figure 7.** Detailed block diagram.

In Histogram Imaging Mode (HIM) the detector integrates for a specified period and then reads out its entire 2-D array as an image. In HIM, each memory location thus represents a pixel location in the detector 2-D 1024x32 array. When a detector event address is sent from the detector to the C&DH, the acquisition control logic updates the proper pixel address location in memory with a single additional count. Each pixel location in the detector array corresponds to a 16-bit word in the acquisition memory and can hold up to $2^{16}-1$ events before the counter saturates at this maximum value. When a histogram exposure is complete, the memory is read out to the LVDS interface to the spacecraft. A second exposure can then begin using the other acquisition memory bank while the first bank is reading out to the spacecraft. Besides the collection and binning of detector events, histogram mode allows the collection of MCP pulse-height distribution (PHD) data from the detector electronics. This PHD data is collected and binned into a 64-bin histogram that is stored within the first two rows of the detector histogram, in a location where no physical pixel within the detector active area exists (therefore, the PHD data does not interfere with the collected detector data). The same C&DH FPGA that controls the pixel-list "ping-pong" acquisitions also controls the histogram data collection mode.

P-ALICE also includes a mechanism to filter out events from selected areas of the detector. This can be used to suppress 'hot pixels' that could develop in the detector, especially in the pixel list mode, which might otherwise consume a large fraction of the available acquisition memory. This filtering is performed before events are processed in either histogram or pixel list mode. Up to 8 rectangular regions of 32x4 pixels can be defined, in order to suppress any events from the selected regions from being processed (e.g., to avoid detector hot spots). Configuration parameters allow for the placement of these filtered areas at any desired positions within the detector area.

## 4. GROUND CALIBRATION RADIOMETRIC PERFORMANCE

Ground radiometric characterization and absolute calibration of the instrument was performed at SwRI's UV space instrument calibration facility located in SwRI's Space Science and Engineering Division.

The radiometric vacuum chamber consists of a 4-inch diameter off-axis parabolic collimator mirror that is fed by a differentially pumped hollow-cathode UV light source[6] and an Acton Research Corporation VM-502 vacuum monochromator. A variable slit and pinhole assembly at the output of the monochromator (and situated at the focus of the collimator mirror) allowed for point source illumination of the P-ALICE airglow and SOC input apertures.

For these calibrations, P-ALICE was mounted to motorized translation and tip/tilt rotation stages within the vacuum chamber that allowed for instrument motion with respect to the collimated input beam. A set of NIST-calibrated photodiodes (one windowless and one silicon PN photodiode) and an AmpTek channeltron were used to measure the relative and absolute fluxes of the collimated beam for absolute effective area measurements. The chamber pressure during the radiometric calibration data runs were in the $5 \times 10^{-7}$ to $5 \times 10^{-6}$ Torr range. The higher-pressure periods were during operation of the hollow-cathode light source when gas was flowing through the light source.

Radiometric characterization tests included the detector dark count rate, wavelength calibration, spectral and spatial point source function (PSF) vs. wavelength, filled slit spectral resolution, off-axis stray light attenuation, and absolute effective area measurements as a function of wavelength.

### 4.1. Dark Count Rate

The detector dark count rate was measured via a set of pixel-list exposures that totaled 9.1 hours of accumulated exposure time. These datasets allowed us to examine both the temporal and spatial distribution of the dark noise and measure its rate. The time hack repetition rate was set to 0.512 sec. Figure 8 shows the event rate during one 1.4-hour exposure. The average dark count rate over the 9.1-hour total exposure interval varied between 2.35 and 2.46 counts sec$^{-1}$ (0.29-0.31 count cm$^{-2}$ sec$^{-1}$) with Poisson counting statistics ($\sigma \sim \pm 0.67$ counts sec$^{-1}$ averaged over 5.12 second bins). This is nicely well below the dark rate specification of <1 count cm$^{-2}$ sec$^{-1}$. Figure 9 shows the accumulated 9.1-hour dark exposure in 2-D histogram format; Figure 10 shows the row and column sums of the total dark exposure. The dark events are uniformly distributed over the entire array with three slightly more active dark regions at (i) center right, (ii) top center, and (iii) bottom center (max rate $\sim 3 \times 10^{-4}$ counts sec$^{-1}$ pixel$^{-1}$). Event pileup at the left and right edges of the array is a characteristic of the DDL detector/electronics, and is evident in Figures 9 and 10.

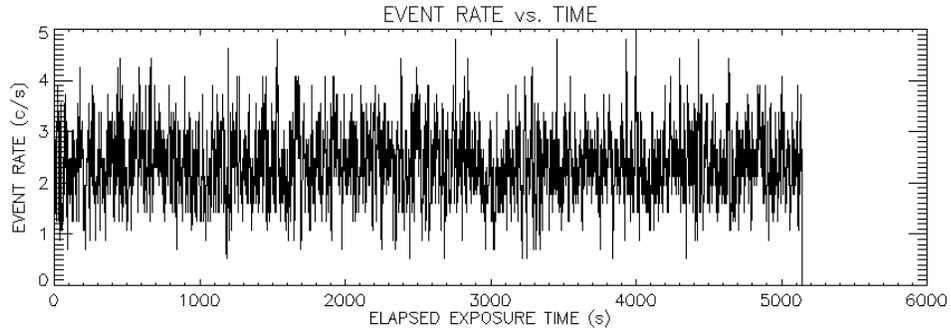

**Figure 8.** Detector dark count rate as a function of time (smoothed over a 5.12 sec period) during one set of dark calibration pixel list exposures totaling ~1.4 hours. The average rate in this exposure was $2.37 \pm 0.67$ counts sec$^{-1}$.

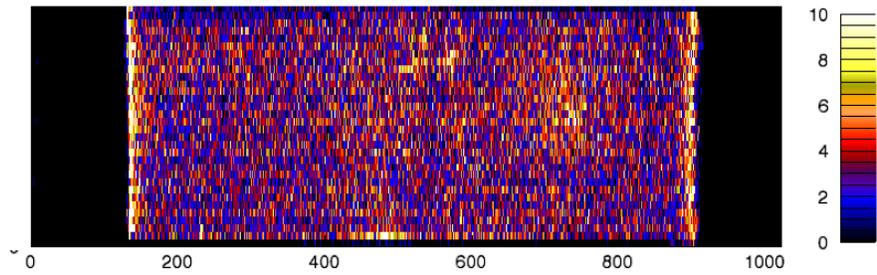

**Figure 9.** Image histogram of the total, 9.1 hour, accumulated dark count exposure. The horizontal axis is the spectral axis (780 active pixels); the vertical axis is the spatial axis (30 active pixels). The vertical color bar scale at right is in units of accumulated counts per pixel over the entire 9.1-hour exposure.

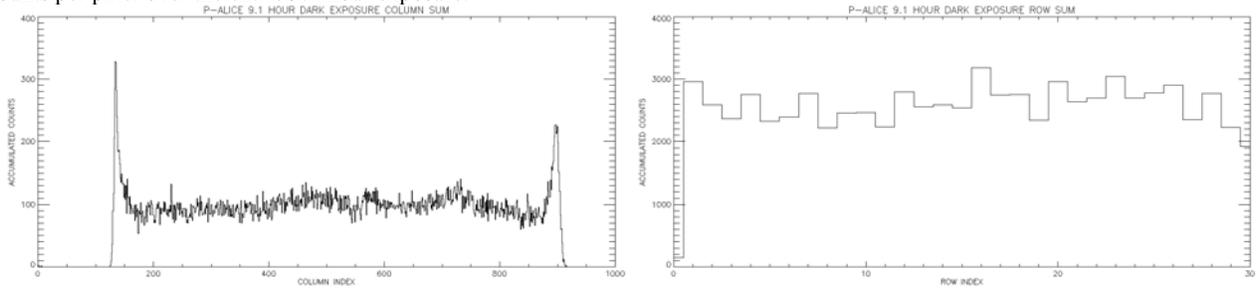

**Figure 10.** (Left) Column sum of the totaled 9.1-hours of dark exposure cal. Note the uniform distribution of events across the width of the detector active area. (Right) Row sum of the total 9.1-hour dark exposure. Note that a few rows (every third row to left and right of center) show slightly more output than expected with just pure spatially uniform noise characteristics (indicative of a pure Poissonian process). This periodic noise is a characteristic of the flat field performance of the detector and may be attributed to "DNL" noise in the analog-to-digital converters in the detector electronics.

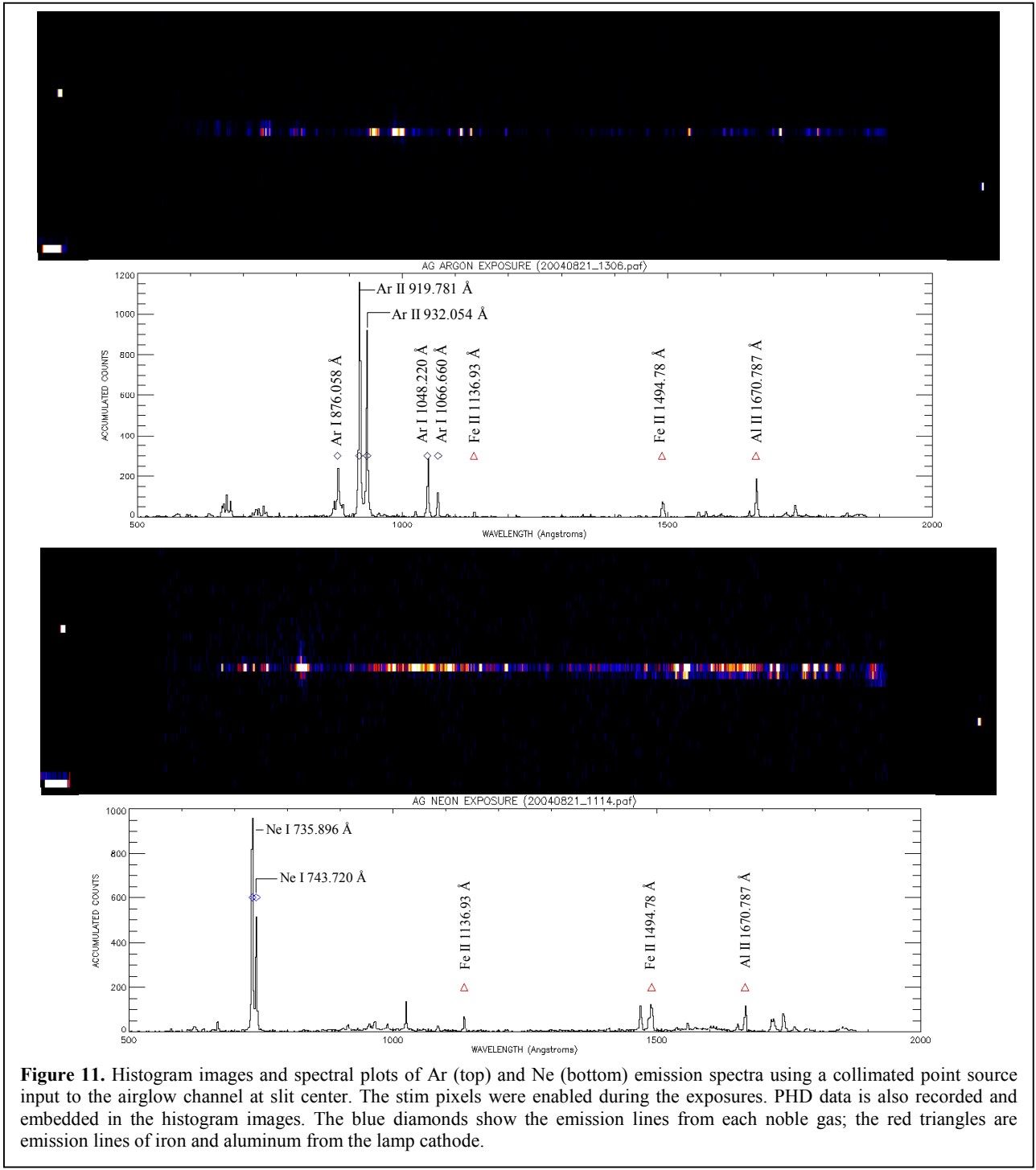

**Figure 11.** Histogram images and spectral plots of Ar (top) and Ne (bottom) emission spectra using a collimated point source input to the airglow channel at slit center. The stim pixels were enabled during the exposures. PHD data is also recorded and embedded in the histogram images. The blue diamonds show the emission lines from each noble gas; the red triangles are emission lines of iron and aluminum from the lamp cathode.

## 4.2. Spatial/Spectral Resolution

### 4.2.1. Airglow Channel Point Spread Function

The spectral and spatial PSFs were measured with the airglow aperture at 0.5° intervals along the length of the slit and at specific wavelengths across the P-ALICE passband using argon and neon gasses in the hollow-cathode UV light source. Point source images were acquired in histogram format at each slit location beginning at the center of the slit (spatial offset of 0°) and at offset angles of ±0.5°, ±1.0°, ±1.5°, ±2.0°, ±2.5°, and ±3.0° with respect to slit center. Figure 11 shows two histogram image exposures showing the recorded emission lines of Ar and Ne at slit center across the P-ALICE passband with the Acton monochromator set at zero order. The two stim pixels and the embedded pulse-height distribution data are also visible in the images.

For each exposure, the image row with the maximum number of counts was used to fit a series of Gaussians for each identified emission line in the spectral axis. In the spatial axis each column was fit with a Gaussian to determine the spatial PSF. Figure 12 shows the spectral PSF as a function of wavelength at slit center using the Ar and Ne exposures. The measured values varied between 3 and 4.5 Å, which satisfied the P-ALICE specification at slit center of <6 Å FWHM. The measured spatial PSF across the P-ALICE passband was <1.6 spatial pixels FWHM, satisfying the requirement of a spatial PSF of <2 spatial pixels FWHM. The spectral and spatial PSFs were very similar in size and shape (even though -2° is in the narrow slit segment and +2° is in the wide slit segment), and varied between 4.4 and 6.5 Å FWHM, with an average of ~5 Å FWHM across the passband. The spatial PSF remained <1.6 spatial pixels FWHM, but grew an average of ~0.2 spatial pixels as the off axis angle was increased.

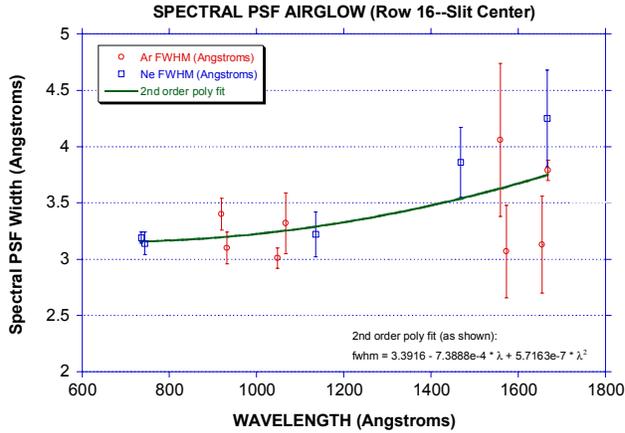

**Figure 12.** A plot of the measured airglow aperture spectral PSF as a function of wavelength at the center of the P-ALICE slit; error bars are 1σ errors based on Gaussian fits of the spectral profiles. The solid line is a 2$^{nd}$ order polynomial fit to the data. Lower count rates at longer wavelength emission lines produce more PSF scatter.

### 4.2.2. Airglow Channel Filled Slit Resolution

Filled slit images as in Figure 13 using a deuterium ($D_2$) lamp to flood the airglow entrance aperture were used to determine the filled slit spectral resolution and the spatial plate scale. A few discrete spectral emission lines exist about the H Lyα emission at 1216 Å (i.e., emission lines are evident on either side of the photocathode gap), and a continuum of emission appears at the right side of the active area. The histogram clearly shows the image of the slit with the wide 2°x2° SOC FOV at the top (the large square pattern) followed by the narrower 0.1°x4° airglow FOV just underneath the SOC FOV (refer to the slit design in Fig. 3). In addition, the expected "slit curvature" aberration is clearly evident. This aberration was well characterized and implies that the wavelength scale offset is a function of the spatial location along the length of the slit.

One isolated emission line just to the left of the gap was used to compute an upper bound to the filled slit resolution. Figure 14 shows this emission at the center of the slit along with a Gaussian fit to it and its companion. The narrow emission has a FWHM spectral width of 9.0±1.4 Å. This width is within the filled-slit resolution requirement of <18 Å FWHM for the airglow channel. The spatial plate scale was also derived from this data and was found to be 0.27°±0.01° per spatial pixel, which meets our <0.3° per spatial pixel requirement. The filled slit spectral resolution in the wide portion of the slit was measured to be 172 Å. FWHM.

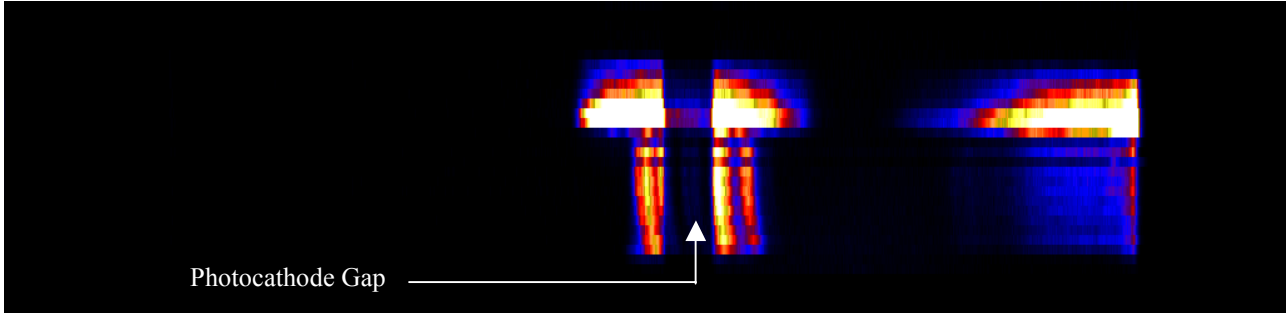

**Figure 13.** A histogram image acquired using a $D_2$ light source flooding P-ALICE's input airglow aperture.

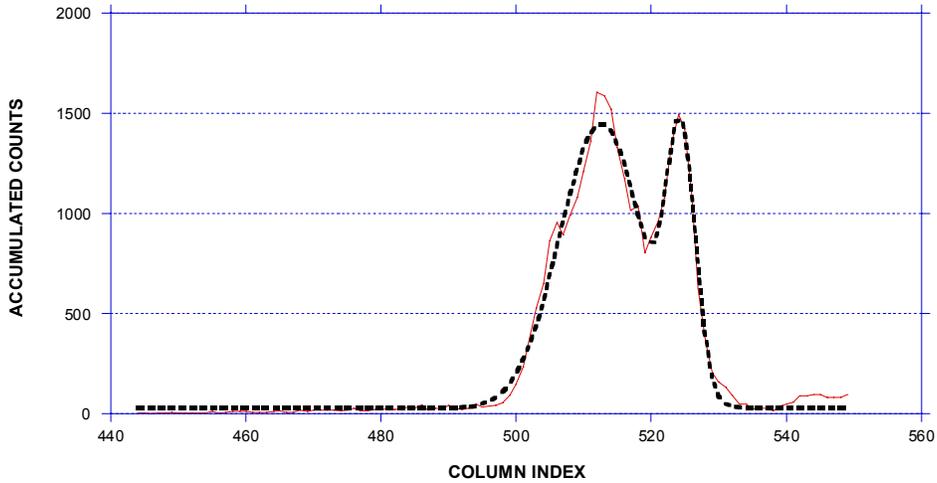

**Figure 14.** Plot of row 16 in the $D_2$ spectral image shown in Figure 13 (solid curve). A two Gaussian sum plus DC offset was fit to this profile (dashed line). The narrow emission on the right was used to calculate the filled slit spectral resolution.

### 4.2.3. Solar Occultation Channel Point Spread Function

As shown in Figure 15, the SOC spectral and spatial PSFs were measured at the center of the 2°x2° slit SOC opening using argon and neon gasses in the hollow-cathode UV light source. The image row with the maximum number of counts was used to fit a series of Gaussians for each identified emission line in these images. The measured spectral PSF as a function of wavelength varied between 2.7 and 3.5 Å FWHM (see Fig. 12), easily meeting the instrument's <6 Å FWHM specification.

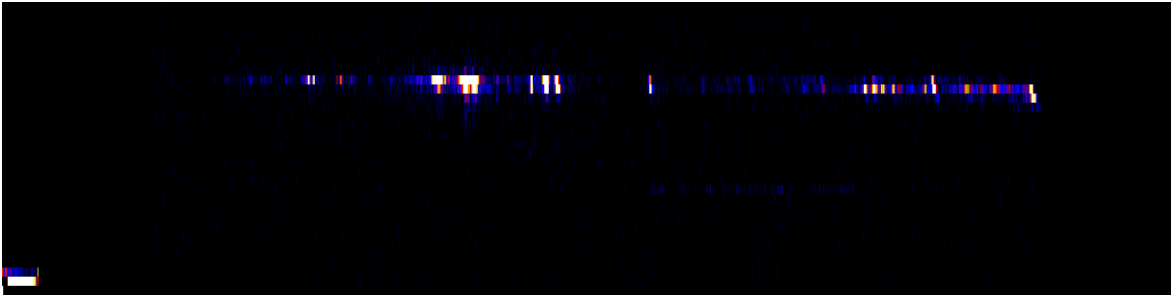

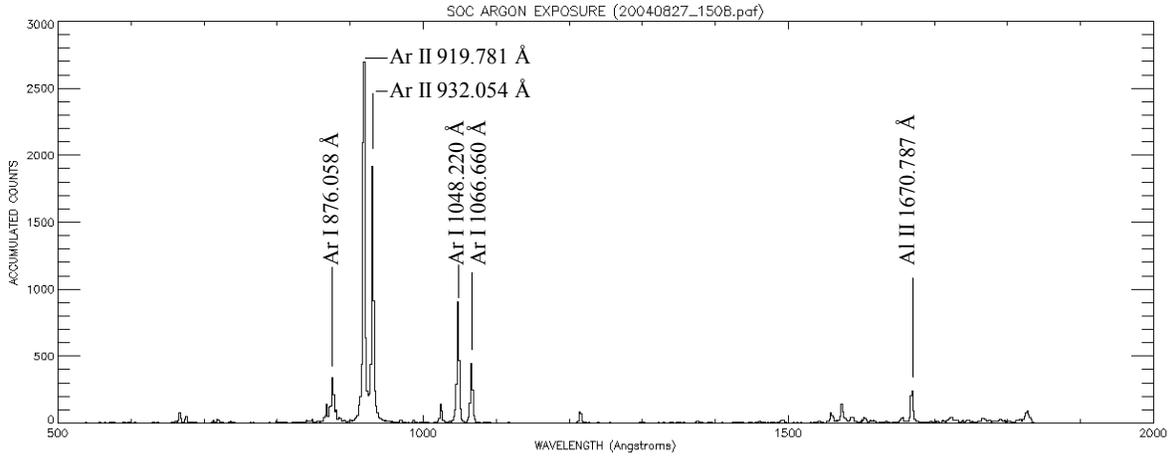

**Figure 15.** (Top) Histogram image of the argon emission spectrum through the SOC. (Bottom) Spectrum of the above histogram image summed across rows 21-23.

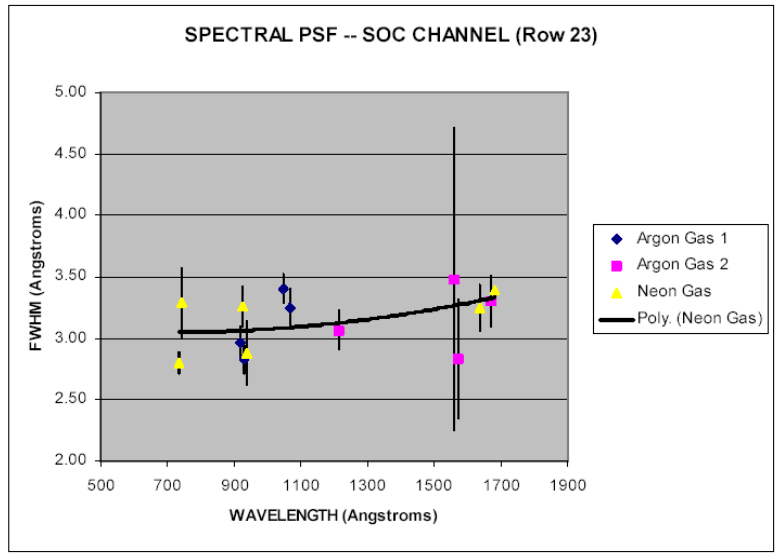

**Figure 16.** The measured spectral PSF at the center of the SOC field of view opening in the slit; error bars are 1σ errors based on Gaussian fits of the spectral profiles. The solid line is a 2$^{nd}$ order polynomial fit to the PSF data.

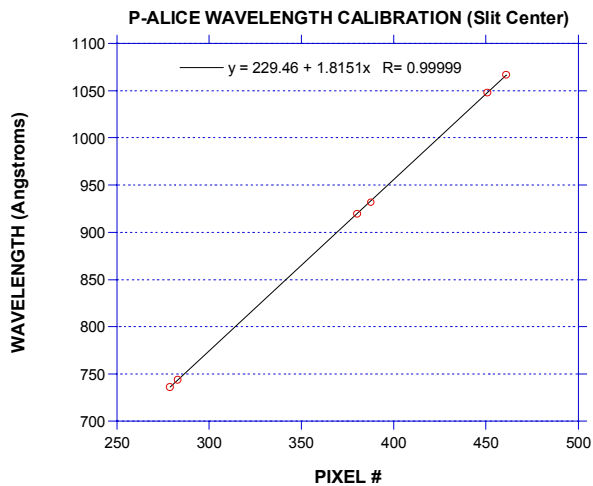

**Figure 17.** Measured positions of discrete Ar and Ne emissions versus wavelength at slit center. The solid line shows a linear fit to the data to determine the spectral plate scale and offset.

### 4.3. Wavelength Calibration

Image histograms taken to determine the airglow PSF values were also used to determine the wavelength calibration at room temperature (+22ºC). Figure 17 shows the pixel locations of the four principal Ar emission lines at 919.781, 932.054, 1048.22, and 1066.66 Å and two bright Ne emission lines at 735.896 and 743.72 Å at slit center. The spectral plate scale and linear offset were determined with a simple linear fit to this data: $d\lambda/dx$=1.815±0.004 Å/pixel; offset (slit center)=229.5±1.5 Å (at pixel=0). The linear fit is good—the $\chi^2$ statistic for the fit is 1.93 with a linear correlation coefficient of 0.99999. The standard deviation of the wavelength residual between the linear fit and the six absolute emission line wavelengths is ±0.62 Å.

The wavelength passband was computed with the above measured plate scale and offset values. The active area starts at spectral pixel 130 and ends at pixel 910, which corresponds to a total wavelength passband of 465-1881 Å, which more than satisfies our passband requirement of 520-1870 Å.

The wavelength offset varies slightly with the temperature of the detector electronics. Therefore, care must be taken to apply the proper offset according to the detector electronics temperature. This ~0.1 pixel (deg C)$^{-1}$ towards longer wavelengths[*] dependence was measured during detector subsystem and instrument thermal vacuum testing. The wavelength offset also varies with the (x, y)-location of the point source image within the slit due to both instrument pointing and to slit curvature aberrations.

### 4.4. Scattered Light Characteristics

#### 4.4.1. Off-Axis Light Scatter

The off-axis light scattering characteristics of the airglow channel were measured in both the horizontal and vertical axes of P-ALICE. Histogram images using a collimated input source and a vacuum UV hollow-cathode lamp gas mixture of H/He were acquired at angular input angles that varied between -8.9° and +9.2° with respect to the airglow boresight axis in the horizontal plane (perpendicular to the slit length with the vertical axis fixed at the center of the slit), and at input angles between -2.5° to +5.0° in the vertical plane (parallel to the slit length with the horizontal axis fixed at the center of the slit)[**].

The acquired histogram images were analyzed to determine the point source transmittance (PST) as a function of incident input angle to the P-ALICE boresight. The PST is defined as follows: PST=$E_{FP}/E_{input}$, where $E_{FP}$ is the irradiance at the focal plane; and $E_{input}$ is the input irradiance at the entrance aperture of the instrument. This expression can be written in terms of the measured off-axis angle count rate to the on-axis count rate ratio ($R_{off}/R_0$); the effective area averaged over the P-ALICE passband ($\overline{A_{eff}}$); the focal plane active area ($A_{FP}$); the average in-band QE of the detector photocathodes ($\overline{QE}$); and the ratio of the input beam area ($A_{beam}$) to the geometric area ($A_g$) of the airglow entrance aperture (since the input beam under fills the entrance aperture). Notice that the PST is not normalized to unity at the zero degree offset position (on axis).

Using these parameters, the PST equation becomes:

$$PST = \left(\frac{\overline{A_{eff}}}{A_{FP}\overline{QE}}\right)\left(\frac{A_{beam}}{A_g}\right)\left(\frac{R_{off}}{R_0}\right). \tag{1}$$

The average effective area and QE values used in Equation 2 to compute the PST were values measured during instrument radiometric tests and detector QE tests at the detector vendor (Sensor Sciences, LLC), respectively: $\overline{A_{eff}}$=0.13 cm$^2$; $\overline{QE}$=0.27; $A_{beam}/A_g$=0.05; $A_{FP}$=7.5 cm$^2$. With these input values, Equation 2 gives a PST 0f 0.003 at an off-axis angle of zero degrees (i.e., on axis along the boresight with $R_{off}/R_0$=1).

---

[*] The wavelength offset value, therefore, shifts towards the blue at a rate of –0.2 Å (deg C)$^{-1}$. The plate scale variation with temperature is negligible and can be ignored.

[**] Volume constraints within the radiometric vacuum chamber limited the input offset angles to these values.

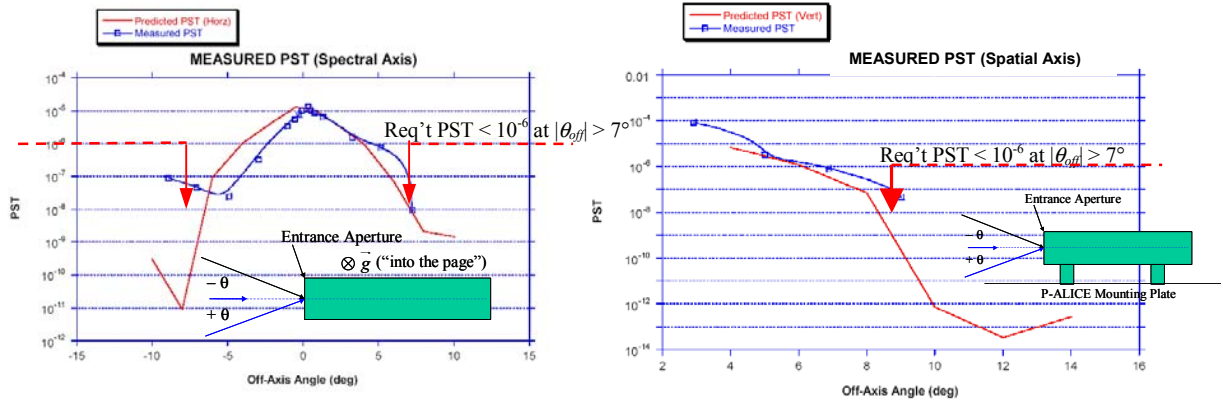

**Figure 18.** The measured PST as a function of the input off-axis angle with respect to the airglow boresight in the spectral axis (left) and spatial axis (right). Both the measured data values (blue squares) and the predicted PST (red solid curve based on the P-ALICE stray light analysis) are shown along with the specification requirement. The PST values near the 0 degree off-axis angle in the spectral axis are all outside the FOV of the slit in the horizontal axis (closest off-axis angle shown is at an off-axis angle of –0.07º where PST=1.0x10$^{-5}$). The "inset" diagrams define the sign of the input offset angle.

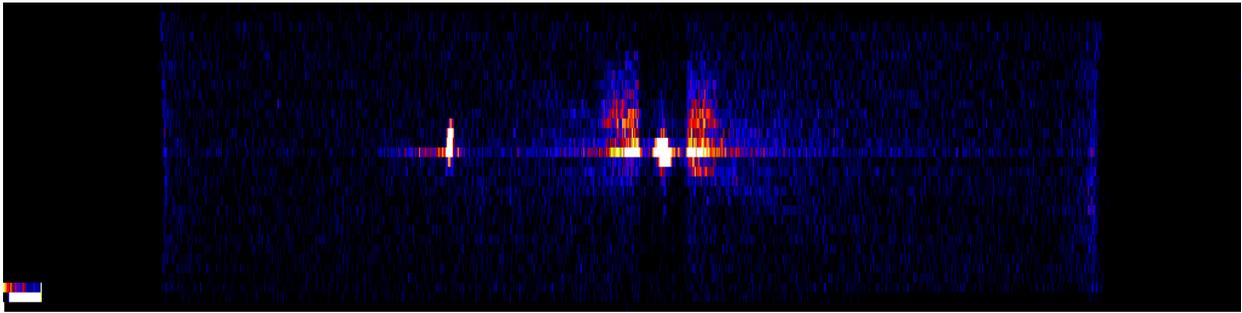

**Figure 19.** Histogram exposure of the H Ly$\alpha$ emission. The primary emission is centered in the photocathode gap near the center of the active area; but the wings of the emission and scattered light from the emission are evident outside the gap. The fainter emission to the left of center is the Ly$\alpha$ grating ghost discussed in the text.

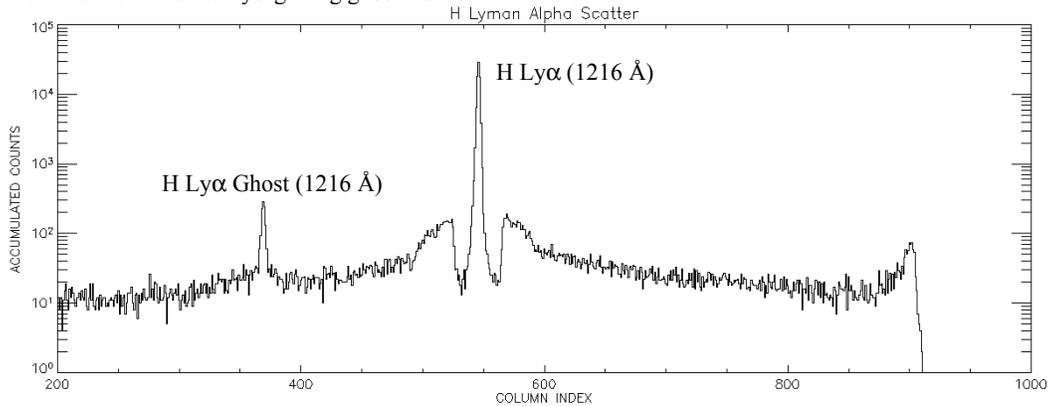

**Figure 20.** Spectral profile of the above histogram image showing the H Ly$\alpha$ emission line centered in the PC gap and the Ly$\alpha$ ghost at 893 Å.

Figure 18 shows the measured PST as a function of off-axis angle in the spectral and spatial axes. Our requirement of a PST <10$^{-6}$ at off-axis angles > 7° was met in both axes. This requirement improves the viability of night-side observations of Pluto and Charon during the flyby.

### 4.4.2. H Lyα Scatter

The total integrated scatter (TIS) at H Lyα was measured using histogram exposures and a monochromatic collimated input light source at 1216 Å. This scatter is due primarily to imperfections in the grating surface. The TIS is defined as the ratio of detected counts outside the photocathode (PC) gap to the total counts within the emission line inside the gap (after normalizing the inside rate with the ratio of the QE outside to inside the gap). This scatter can be expressed as follows:

$$TIS = \frac{R_o}{R_t}, \quad (2)$$

where $R_o$ is the rate outside the PC gap; and $R_t$ is the total normalized detector rate[*]. This ratio can be expressed in terms of the two measurable values $r_t$ (the total detector rate) and $r_i$ (the measured rate inside the gap). Equation 2 reduces to

$$TIS = \frac{r_t - r_i}{(k-1)r_i + r_t}, \quad (3)$$

where $k$ is the ratio of the photocathode QE to that of the bare MCP glass inside the gap. Using the measured ratio of effective area just outside the PC gap to that inside the gap (this ratio is nearly equivalent to the ratio of QE values), we find that k=9.2. Inserting the above values into Equation 3 gives a TIS for H Lyα of 0.024 (2.4%). This result is well within the TIS goal of <10%.

Figures 19 and 20 show, respectively, the 300 sec histogram exposure and spectral profile taken to make this measurement. The H Lyα emission line within the PC gap is evident along with the wings of the emission line protruding on either side of the PC gap, as well as light scatter across the active area. In addition, a ghost image of the Lyα line shows up left of center in the image. This ghost was also noted in the R-ALICE instrument. Ray trace studies of P-ALICE show that this ghost is caused by the reflection of the H Lyα emission line off the front surface of the detector MCP back towards the grating, which then re-diffracts the light back onto the detector surface. The intensity of this ghost is nearly identical to that measured with R-ALICE.

### 4.5. Absolute Effective Area

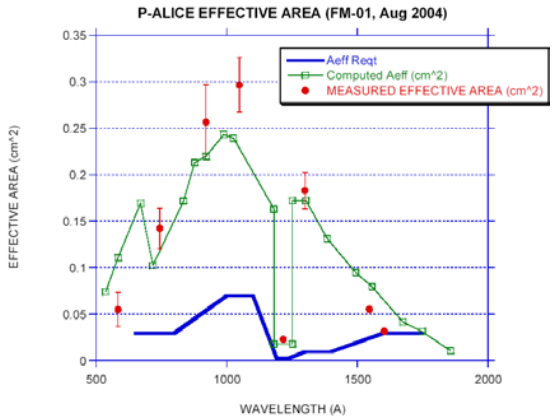

**Figure 21.** The measured effective area for the airglow channel (red solid circles) plotted versus wavelength. Also shown are the computed effective area estimates (green open squares) based on previously measured SiC reflectivity values of the OAP mirror and grating and detector QE values. The effective area specification is shown as the solid blue line in the lower portion of the graph.

#### 4.5.1. Airglow Channel

The effective area of the airglow channel was measured across the P-ALICE passband at discrete wavelengths using emission lines from neon, argon, and a hydrogen/helium gas mix covering a wavelength range of 584–1603 Å. Histogram exposures were made with the chamber's monochromator set to the specific measurement wavelength. The monochromator output slit was set to provide a detector output count rate that did not show any saturation effects in the pulse-height distribution (i.e., too high a local flux on the MCP stack will cause a local gain drop that is evident in the PHD data). Background exposures were also taken for each illuminated exposure to allow for dark subtraction.

The beam flux was measured for each effective area measurement using a NIST calibrated silicon (Si) photodiode. In most cases, the beam flux had to be increased in intensity to allow enough flux for a measurement with the photodiode. However, this increased level flux was too high for a measurement with P-ALICE causing local MCP saturation. An Amptektron channeltron was used to measure the ratio of the beam fluxes at low and high intensity levels to provide a correction factor to the flux measured by the NIST photodiode.

---

[*] These detector rates have been corrected for electronic deadtime effects.

Figure 21 shows the measured effective area as a function of wavelength $\lambda$, $A_{eff}(\lambda)$. The equation used to compute $A_{eff}$ is

$$A_{eff}(\lambda) = \frac{A_g R(\lambda)}{\Phi(\lambda)}, \qquad (4)$$

where $A_g$ is the geometric area of the P-ALICE airglow entrance aperture (16 cm$^2$); $R(\lambda)$ is the detector count rate (corrected for the detector deadtime and the detector background), and $\Phi(\lambda)$ is the input flux to the instrument based on measurements with the NIST Si photodiode and the channeltron. The red circles in the figure show the measured effective area values; the green squares show the predicted values based on reflectivity values of the OAP mirror and grating (including the grating diffraction efficiency), and the photocathode QE response measured during subsystem level optics and detector tests. The blue solid line (no data points) is the minimum effective area requirement set for the instrument. As is evident in Figure 21, the effective area requirement was met with significant margin.

### 4.5.2. Solar Occultation Channel

The effective area of the SOC was measured, and the relative sensitivity compared with the airglow channel was computed. The airglow channel is 3000-7300 times more sensitive than the SOC channel (depending on the wavelength), with an average ratio of ~5000. The ratio of airglow to SOC effective area requirement is set to limit the detector count rate to <30 kHz during the solar occultation observations at Pluto/Charon. This ratio must be >2400 to meet this requirement; this requirement was easily met.

**Table 1—Summary of P-ALICE Ground Radiometric Test Performance**

| Radiometric Test | Test Requirement/Goal | Measured Values |
|---|---|---|
| Dark count rate | ≤1 cm$^{-2}$ sec$^{-1}$ (total array) | 0.3 cm$^{-2}$ sec$^{-1}$ (total array) |
| Wavelength calibration | <1 Å calibration precision | ±0.62 Å (1$\sigma$ using a linear fit) |
| PSF vs. $\lambda$ (airglow & SOC) | <6 Å (SOC)<br><18 Å (airglow)<br><2 spatial pixels (spatial) | 3.0-3.5 Å (SOC)<br>3.0-4.5 Å (airglow)<br>0.4-1.5 spatial pixels (spatial) |
| Wavelength passband | 520-1870 Å (minimum coverage) | 465-1881 Å |
| Filled slit spectral resolution | <18 Å FWHM (airglow) | 9.0±1.4 Å (FWHM) |
| Spectral plate scale | 1.7±0.2 Å/pixel | 1.832±0.003 Å/pixel |
| Spatial plate scale | <0.30 per spatial pixel | 0.27°±0.01° |
| Off-axis light scatter test (airglow) | PST* <10$^{-6}$ at $\theta_{off}$>7° | PST <4.6x10$^{-8}$ at $\theta_{off}$>7° (spectral axis)<br>PST <8.3x10$^{-7}$ at $\theta_{off}$>7° (spatial axis) |
| H Ly$\alpha$ attenuation/scatter | <10% (total integrated scatter) | TIS=2.4% (outside PC gap) |
| H Ly$\alpha$ gap $\lambda$ boundaries | $\Delta\lambda$~70-75 Å centered at 1216 Å | Gap: 1178.6 Å to 1251.9 Å ($\Delta\lambda$=73.3 Å)† |
| Absolute effective area (airglow aperture) | 650-800 Å : >0.03 cm$^2$<br>1000-1100 Å : >0.07 cm$^2$<br>1181-1251 Å: 0.003-0.03 cm$^2$<br>1300-1400 Å : >0.01 cm$^2$<br>1600-1750 Å : >0.03 cm$^2$ | 650-800 Å : 0.07-0.17 cm$^2$<br>1000-1100 Å : 0.20-0.30 cm$^2$<br>1181-1251 Å: 0.02 cm$^2$<br>1300-1400 Å : 0.15-0.18 cm$^2$<br>1600-1750 Å : 0.03-0.05 cm$^2$ |
| SOC relative effective area | <1/2500 of AGC effective area** | 1/7300 to 1/3000 of AGC effective area |

*PST=Point Source Transmittance=$E_{det}/E_{in}$, where $E_{det}$ is the irradiance at the detector; $E_{in}$ is the input irradiance. **Assumes a nominal AGC effective area. †With point source centered in AGC slit; gap pixel boundaries: cols 525.5 to 565.4 at room temperature,+22° C).

## 5. IN-FLIGHT PERFORMANCE

### 5.1. Commissioning Test Overview

In-flight commissioning of P-ALICE began about one month after launch of NH in February 2006. The first set of commissioning tests consisted of the first in-flight power on of the instrument followed by a set of sequences that tested the program memory and command and telemetry interfaces with the spacecraft. In addition, the aperture door launch latch was fired open and the door exercised to verify its function. These early tests while the spacecraft was still outgassing did not involve detector high voltage.

A second set of commissioning tests were run in May 2006 that included sequences to test the optics heaters for decontamination control, to open the detector door, and a ramp up of the detector HVPS to full operational voltage. These activities were followed by sequences to obtain the first-light exposures of the H I Lyman-α sky glow and exposures of the dark background. An additional instrument-commissioning period to calibrate the in-flight sensitivity performance of P-ALICE using UV stars took place in September 2006.

Performance in the March 2006 set of commissioning tests were entirely nominal. These tests occurred with P-ALICE at room temperature (~20°C); the light-time distance of NH from Earth was between 112 and 115 seconds. The measured power draw of P-ALICE was nominal at ~3.6 W, and all the initial mode commanding and memory/C&DH tests showed proper performance. The heaters on the OAP mirror, grating and SOC mirrors were activated and showed nominal heating profiles; in addition, the detector stim pixels were activated and exposures taken that verified proper stim pixel locations and stim widths. Finally, the aperture door launch latch was successfully fired open, and the door was exercised to verify operation of the door LAT motor. At the conclusion of these first commissioning tests, P-ALICE was powered off until early May, when the second set of tests were run.

The May 2006 set of commissioning tests began with initial power on of the instrument followed by mode checks, stim pixel exposures, and two 24-hour long optical heater decontamination sessions. Performance in all of these tests was nominal The one-way light time distance was ~25 minutes for these operations. The detector door was successfully opened to allow the interior of the detector housing to outgas to space. Twenty-four hours after opening the detector door, the HVPSs were activated to a first step of –2.5 kV, and tests to verify the on-board safety triggers were successfully run (i.e. HV and MCP strip current values as well as excessive detector count rates). A noise immunity test was also performed that showed no anomalous behavior of the instrument electronics with HV on. Following the –2.5 kV HV level test, the HVPS was then successfully raised to the full operational level of –4.5 kV, again with no anomalous behavior. HV was then powered off and a 12 hour optical decontamination heater session ensued. Following the decontamination session, the HV was powered to –4.5 kV, and a set of dark and sky glow images were obtained in both pixel list and histogram modes. The detailed results of dark and sky glow exposures are discussed next.

## 5.2. Dark Count Rate

Two 600-second histogram exposures were taken with the aperture door closed and the detector HV set to –4.5 kV. The detector electronics temperature was 21.9°C and the detector housing temperature was 19.6°C. Figure 22 shows the image histogram of both exposures co-added together; Figure 23 shows the row and column sums of this dark exposure.

The stim pixels were activated during these exposures and are shown in the Figures 22 and 23. The average detector count rate of the co-added exposure was 98.5 counts per second after subtraction of the stim pixel rate of 18 c/sec. This background rate includes contributions from 1) the intrinsic MCP dark noise (typically 1-5 c/sec as measured during ground vacuum tests); 2) atomic particles in the local environment (protons, electrons, other ions); and 3) radiation from the spacecraft's RTG (neutrons and gamma-rays). Assuming a similar combined MCP intrinsic dark rate and ambient space background rate to that observed with Rosetta-ALICE of ~20 c/sec, the contribution from the RTG is ~80 c/sec, due mostly to gamma-rays. This RTG background rate is within our predictions based on previous RTG exposure calculations and tests with the ALICE detector during the NH integration and test program.

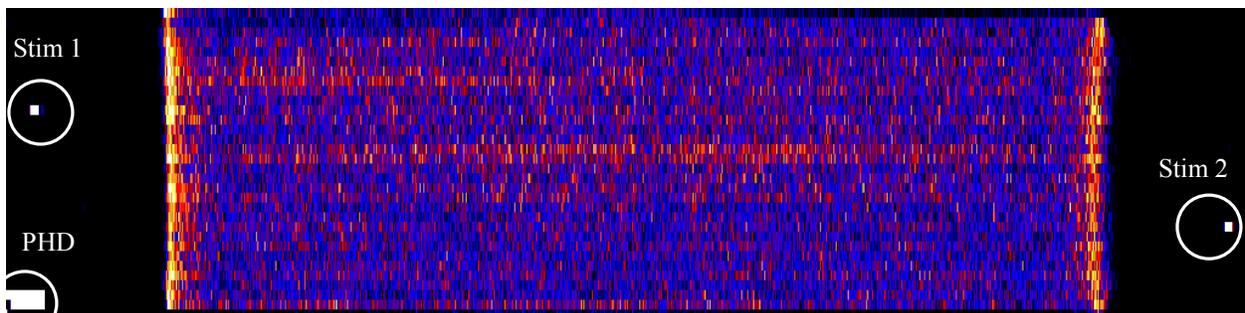

Figure 22. Detector background (dark) histogram image made with the ALICE aperture door shut after the detector door was opened. The two stim pixel locations and the PHD data are highlighted in the image. The higher row amplitudes are due primarily to DNL electronic noise caused by the high gain pulse contribution in the dark background pulse-height distribution.

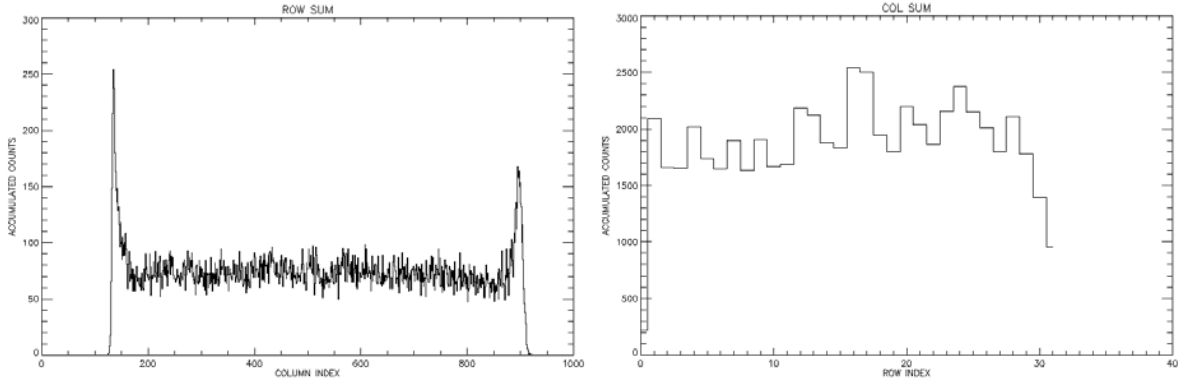
Figure 23. The row (left) and column (right) sum of events of the first background exposure.

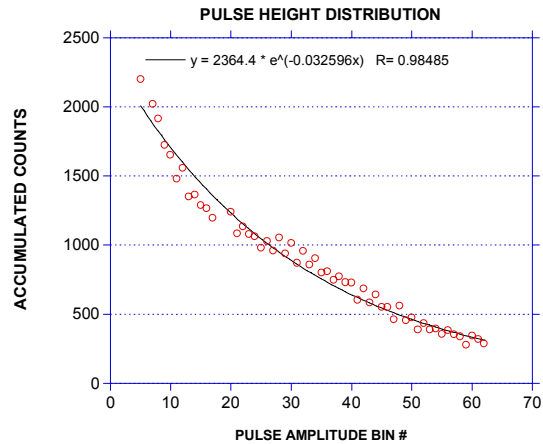
Figure 24. The PHD of the dark background histogram exposures showing its negative exponential shape. The black solid line through the data points is a negative exponential fit to the data.

Figure 24 shows the recorded PHD of the background exposures obtained in May 2006. The background events contribute to the negative exponential shape of the curve—this is the expected shape caused by gamma-ray events being converted to electrons throughout the entire thickness of the MCP stack.

### 5.3. First Light Images

At the conclusion of the background exposures, P-ALICE was commanded to take histogram and pixel-list exposures of the Lyman-alpha skyglow. P-ALICE's airglow boresight was centered at celestial coordinates (J2000) RA = 107.150° (7h 8m 36s) and DEC = +22.205° (+22° 12' 18"), which is near the center of the constellation Gemini. The ecliptic coordinates of this nominal boresight RA, DEC position is longitude = 105.84°; latitude = -0.3°. Three 600-second histogram exposures and a 60-second pixel list exposure were taken with the detector HV set at –4.5 kV, and the aperture door in the open state.

### 5.3.1. Image Histogram Exposures

Figure 25 shows an image of the three initial sky histogram exposures co-added together. The H Ly-α emission line is the strongest in the spectrum, and the exposure shows a nice image of the slit aperture. Also note that H Ly-β at 1026 Å, H Ly-γ at 972 Å, and He I at 584 Å are seen, along with the known H Lyman-alpha "ghost" at 893 Å[1]. Light at wavelengths >1216 Å are due primarily to scattered Ly-α light off the grating (see Slater et al. 2005), and possibly of a faint UV star within the SOC field-of-view (FOV) (see Figure 25). Figure 26 shows the emission line profiles from the three co-added exposures with the background removed within the narrow airglow slit region only (co-addition of rows 5-17). The best filled-slit spectral resolution occurs in rows 15-17 (where the P-ALICE boresight falls), with FWHM values in the range of 10.0-10.5 Å. This matches closely with the measured value of 9.0±1.4 Å measured during ground radiometric test (see Slater et al. 2005). Note also the scattered light plateau of ~100 counts in Figure 26 from ~900-1800 Å. This translates to a scattered light contribution of ~0.006 c/s per pixel in this region of the focal plane (or ~1% of the total flux from the skyglow flux illuminating the focal plane).

---

[1]The H Lyman-alpha "ghost" is a known feature of P-ALICE caused by the reflection of this emission line off the detector MCP surface back towards the grating, which diffracts it back to the focal plane (see Slater et al. 2005).

The count rates in both the narrow and wide slit segments can be used to estimate the effective area provided we know the brightness of the emission. For the H Lyman-α emission, the estimated Ly-α sky brightness was ~200 Rayleighs (with an estimated error of ±30%) in the direction of the P-ALICE boresight. This emission line fills both the 2° x 2° SOC FOV (top rectangular region in Figure 25), and the 0.1°x4° narrow airglow FOV. At H Ly-α, the narrow slit illuminates only the photocathode gap region which has a detective quantum efficiency one-tenth that of each cathode on either side of the gap. The SOC FOV, on the other hand, fills both the 70-Å wide gap and ~110 Å of photocathode split nearly evenly on either side of the gap. This is clearly evident in the histogram image shown in Figure 25 (top), which shows the two bright emission features on either side of the gap in the top portion of the slit image, and the fainter filled gap region along the stem of the slit image. Note that the H Ly-α emission is slightly "blueward" of the gap center by approximately 7 Å. This is within the measured preflight value of 3± 5Å blueward of the gap center (the uncertainty here is based on the exact position of the point source within the 0.1° wide airglow gap).

For H Ly-α, the count rate in the combined background-subtracted image histogram within the SOC 2° x 2° slit was 2220 c/sec, and within the narrow 0.1°x4.0° slit it was 45 c/sec. This translates to an effective area in the gap region of ~0.02 cm$^2$, and an effective area on either side of the gap of 0.17 cm$^2$. Figure 27 shows the effective area as a function of wavelength measured during ground test; note that the new in-flight measured values are close to the ground measurements. Stellar calibrations taken later provided more exact in-flight effective area values at a variety of wavelengths, but these measurements provided strong encouragement that the instrument was operating to spec in space.

**Table 2. Predicted emission brightnesses from the skyglow exposures assuming the ground-based effective area values. The count rate in column 4 is the net rate after subtracting the dark background over the entire slit FOV (including both the narrow and wide slit segments).**

| Emission Line | | Effective Area (cm^2) | Count Rate (c/s) | Brightness (Rayleighs) |
|---|---|---|---|---|
| H Ly-beta | 1026 A | 0.25 | 41 | 1.5 |
| He I | 584 A | 0.06 | 6 | 1.0 |

Table 2 lists the inferred brightness of the H Lyman-β and He I 584 Å emissions assuming the ground-measured effective area values shown in Figure 27. Note that both emissions are in the 1.0-1.5 Rayleigh brightness level (for He I, this emission strength is comparable to measurements by EUVE—e.g., see Flynn et al. 1998). The brightness of the H Lyman-γ emission was difficult to estimate because of its low SNR of its profile.

### 5.3.2. Pixel List Exposures

A 61-second pixel list integration was taken of the skyglow after completion of the three histogram skyglow exposures discussed above. The time hack interval was 0.004 seconds; the discriminator voltage was 0.5 V. Figure 28 shows the digital count rate as a function of time during the exposure, with each data point the average over 40 milliseconds. The average digital count rate was 2608 c/sec (including stim pulsers, which were on); this gives a standard deviation in the plot of 51 c/sec.

Figure 29 shows a histogram of the number of events per time hack interval, along with a Gaussian fit to the data. The excellent fit to the data (reduced $\chi^2$=4.7) indicates Poissonian counting statistics dominate (as expected for photon signals), and that there are no major contributions in the dataset that are non-Poissonian in nature, as can sometimes result from hot emission points from the MCP stack.

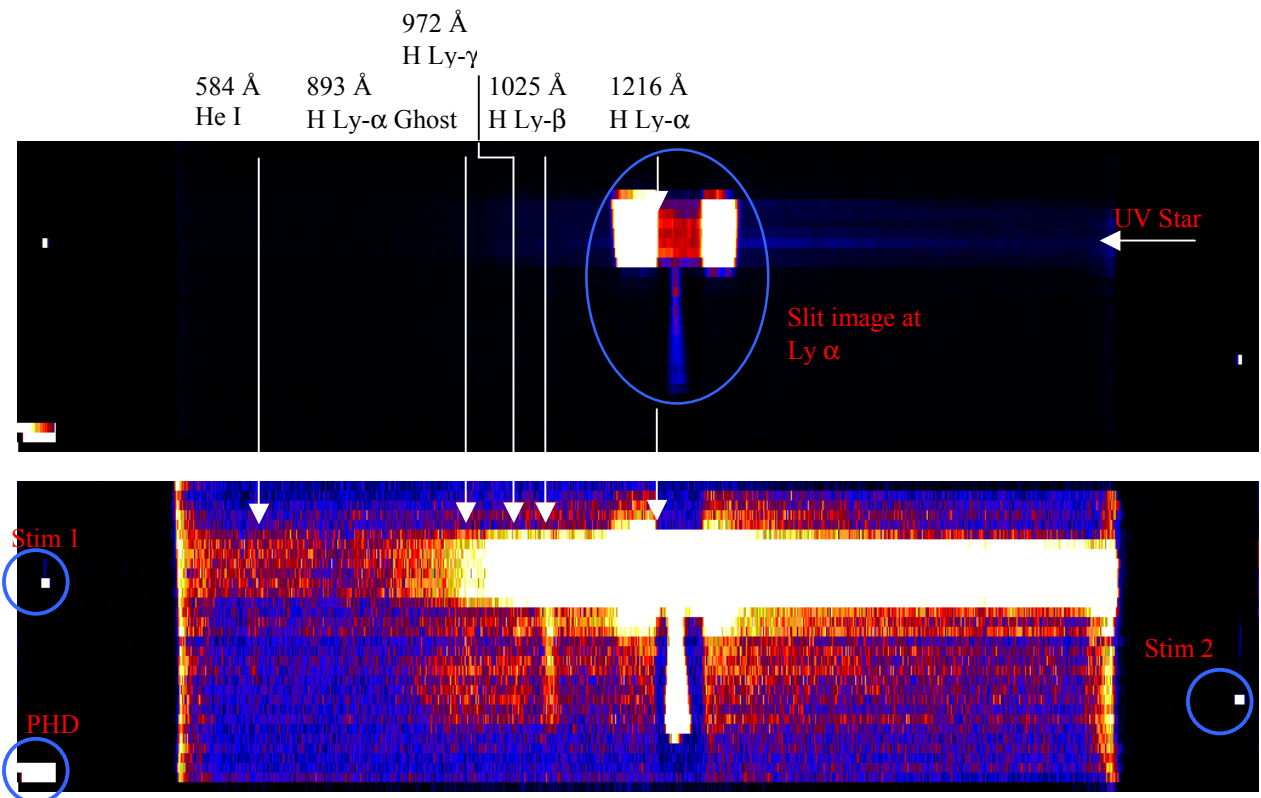

Figure 25. The Lyman-alpha skyglow "first-light" image histogram made up by the co-addition of the three separate 600-second exposures. (Top) Image lookup table set to show off the H Lyman-α emission at 1216 Å. A possible UV star's continuum spectrum also appears in the lower portion of the SOC FOV. (Bottom) A stretched lookup table that shows the fainter emissions at 1026 Å (H Ly-β); 972 Å (H Ly-γ); 584 Å (He I); and the H Ly-α ghost at 893 Å. The two stim pixels and PHD data are also shown.

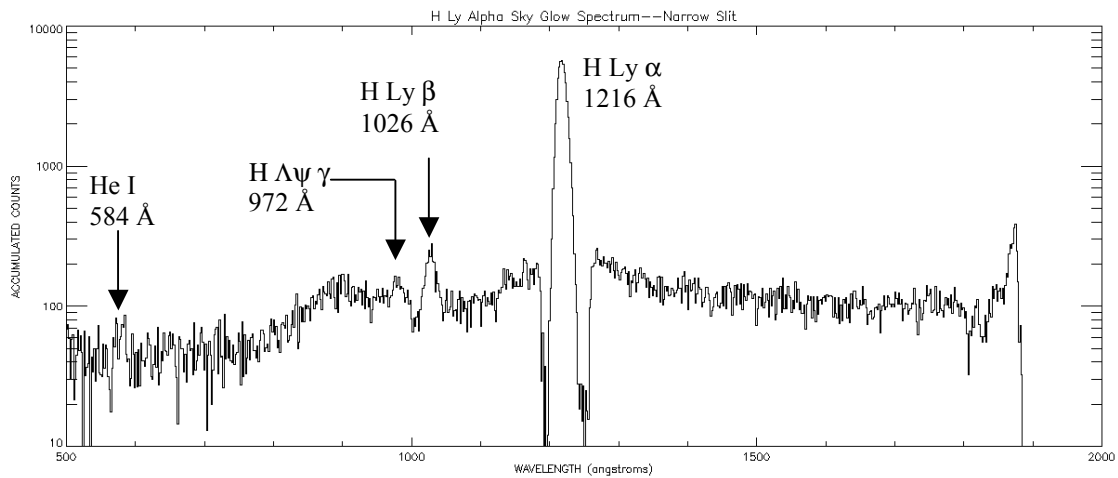

Figure 26. Semi-log plot of the background-subtracted accumulated counts as a function of wavelength accumulated within the narrow airglow slit showing the emissions from H Ly-α (1216 Å); H Ly-β (1026 Å); H Ly-γ (972 Å); and He I (584 Å). Note that the Ly-α emission is within the photocathode gap; hence, the much lower background events immediately surrounding the emission line.

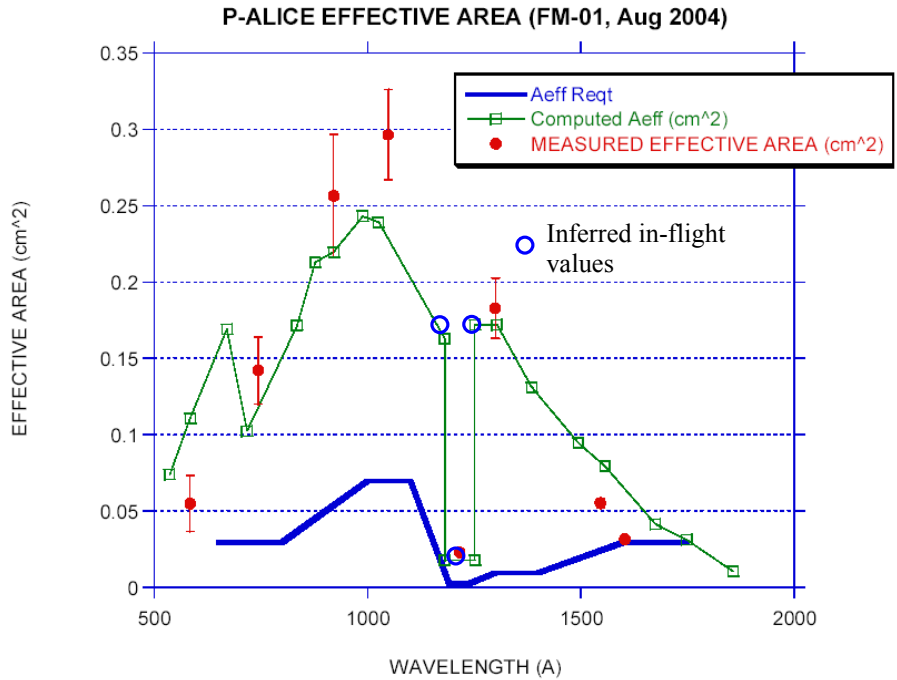

Figure 27. The ground-measured P-ALICE effective area curve with wavelength. Superimposed on this plot are the initially inferred in-flight effective area values within the PC gap and just outside the PC gap (blue open circles) assuming a 200 Rayleigh emission brightness at skyglow H Ly-$\alpha$.

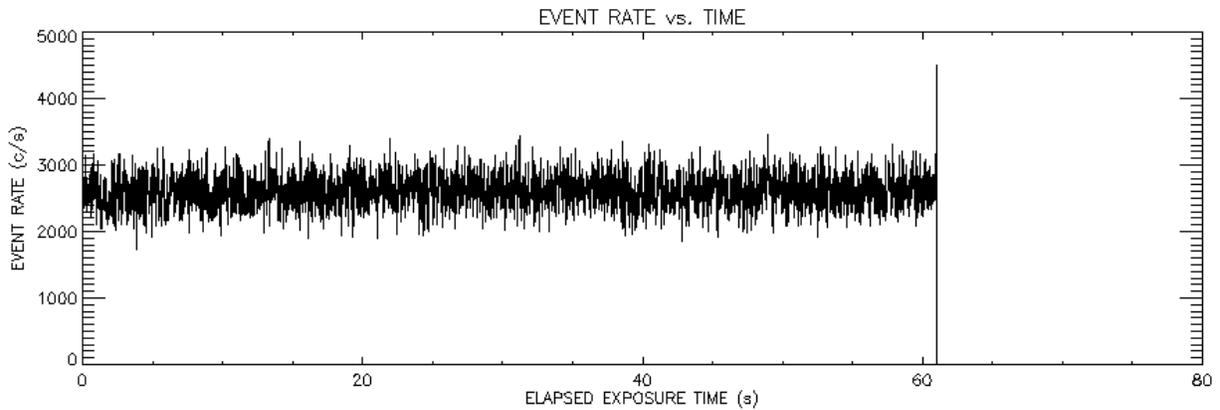

Figure 28. The digital count rate as a function of time during the 61-s pixel list exposures.

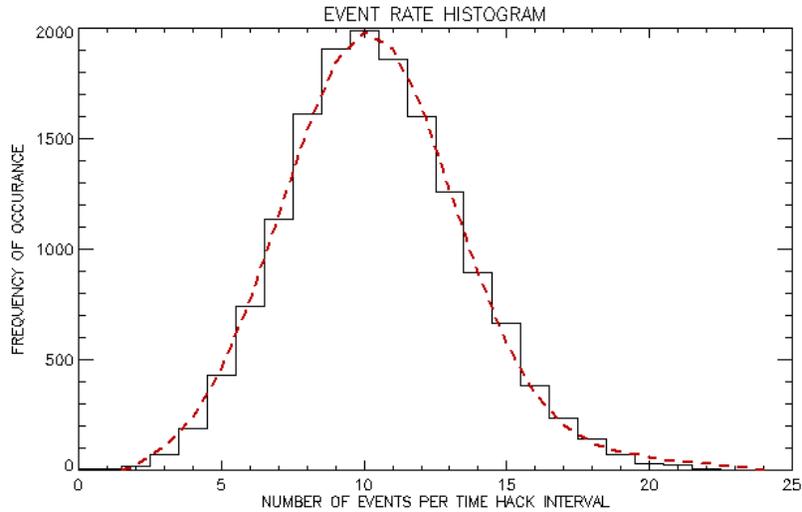

Figure 29. The event rate histogram data (solid line). The red-dashed curve is a Poisson fit to the histogram data.

## 6. Conclusion

The New Horizons ALICE UV spectrometer has been described above. It was successfully launched on 19 January 2006 and is operating normally in space. All in flight performance tests to date have shown performance within specification; the pointing and AGC sensitivity tests completed in September 2006 are in analysis, and the initial results of these tests indicate nominal performance. The primary remaining tests to be performed will be testing of the solar occultation aperture after it is opened near the end of 2006, and instrument mutual noise susceptibility testing in the spring of 2007.

## ACKNOWLEDGEMENTS

We thank the entire P-ALICE engineering and support team at SwRI, as well as John Vallerga and Rick Raffanti of our detector supplier at Sensor Sciences, and our French collaborator Jean-Loup Bertaux for providing optics. This work was supported by NASA contract #NASW02008.